\documentclass{article}
\usepackage{spconf,amsmath,graphicx}
\usepackage{comment,psfrag,amssymb,booktabs,tabularx} % AE

\title{CODING OF DISTORTION-CORRECTED FISHEYE VIDEO SEQUENCES\\USING H.265/HEVC}%
\name{Andrea Eichenseer and Andr\'e Kaup}
\address{Multimedia Communications and Signal Processing\\
	Friedrich-Alexander University Erlangen-N\"urnberg, Cauerstr. 7, 91058 Erlangen, Germany\\
}
\begin{document}
%\ninept % font size 9 instead of 10
%
\maketitle
\begin{abstract}
Images and videos captured by 
%ultra wide-angle cameras such as 
fisheye cameras exhibit strong radial distortions due to their large field of view.
% that makes it possible to project a hemispherical view onto a two-dimensional image plane.
Conventional 
%motion 
%estimation and
%compensation 
intra-frame as well as inter-frame prediction techniques as employed in hybrid video coding schemes are not designed to cope with such distortions, however.
So far, captured fisheye data has been coded and stored without consideration to any loss in efficiency resulting from radial distortion.
%and distortion correction has been applied as a post-processing step.
This paper investigates the effects on the coding efficiency when applying distortion correction as a pre-processing step as opposed to the state-of-the-art method of post-processing.
Both methods make use of the latest video coding standard H.265/HEVC and are compared with regard to objective as well as subjective video quality.
It is shown that a maximum PSNR gain of 1.91~dB for intra-frame and 1.37~dB for inter-frame coding is achieved 
%for the luminance component 
when using the pre-processing method.
Average gains amount to 1.16~dB and 0.95~dB for intra-frame and inter-frame coding, respectively.
%On average, gains of 1.16~dB for intra-frame and 0.95~dB for inter-frame coding are obtained. %mean over LD+RA!!!
%Equivalent maximum and average bitrate savings amount to 31.76~\% (31.47~\%) and 19~\% (21~\%) for intra-frame (inter-frame) coding, respectively.
\begin{comment}
Furthermore, it was found that the radial distortion does not only affect inter-frame prediction, but also intra-frame prediction, letting us conclude that radial distortion compensation for hybrid video coding warrants further investigation.
[TODO; too long!, change last sentence]

[PLACEHOLDER]
Due to its large field of view, wide-angle imagery typically incorporates strong
radial distortions, which conventional intra prediction as well as motion compensation techniques as employed
in hybrid video coding schemes are not designed for.
So far, captured wide-angle data has been coded and stored without considering
any loss in efficiency and distortion correction has been applied as a
post-processing step.
This paper investigates the effects on the coding efficiency when applying the
distortion correction as a pre-processing step.
As expected, gains of up to X in bitrate and Y dB in terms of luminance PSNR can be
achieved, letting us conclude that this alternative processing order is well
worth a consideration. [beginning off-topic, conciseness!]
\end{comment}
\end{abstract}
%

%#####

\section{Introduction}
\label{sec:intro}

In many application scenarios, extreme camera optics that are quite different from conventional rectilinear lenses are made use of.
Such scenarios include surveillance, automotive, or outdoor applications, where typically ultra wide-angle lenses (e.\,g., fisheye lenses) or catadioptric cameras consisting of both lenses and mirrors are employed to capture extensive fields of view of up to and above 180 degrees.
% in both horizontal and vertical direction.
Fields of view of this dimension have the distinct advantage of being able to preserve all the visual content of a hemisphere and map it onto the image plane.
While this is indeed useful for surveillance and other applications, the mapping leads to the introduction of strong radial distortions in the final images and videos. 
This in turn leads to the captured material exhibiting properties that are not taken into consideration by conventional hybrid video codecs such as the recently standardized H.265/HEVC~\cite{hevc,hevc2} or its predecessors.

Hybrid video codecs are block-based and contain procedures for predicting the current block either spatially from the current video frame (intra-frame prediction) or temporally from preceding frames (inter-frame prediction).
When using inter-frame prediction, translational motion between frames can be estimated, compensated, and thus exploited for efficiently coding the current block.
Motion estimation and compensation techniques realizing this inter-frame prediction are typically designed for video material with rectilinear properties, and do not take into account that translational motion in radially distorted video sequences 
%effectively leads to deformations that in turn impair the block-matching procedure.
does not preserve shapes, which in turn impairs the block-matching procedure.
Similarly, intra-frame prediction works very efficiently on images containing clear lines and edges.
As radially distorted images contain arcs instead of lines, efficiency decreases accordingly. 

In an application scenario where it is necessary to correct the radial distortion,
%from the video sequences
the distortion correction is conventionally done after coding.
While the encoding step is typically part of the capturing and storing of the data, distortion correction is a post-processing step after decoding and thus not integrated into the camera.
Considering the above-mentioned inefficiency of 
%motion compensation 
intra-frame and inter-frame prediction
in the presence of radial distortion, it seems prudent to employ the correction as a pre-processing step instead.

\begin{figure}[t]
\begin{minipage}[b]{1.0\linewidth}
  \centering
  \centerline{\includegraphics[width=8.5cm]{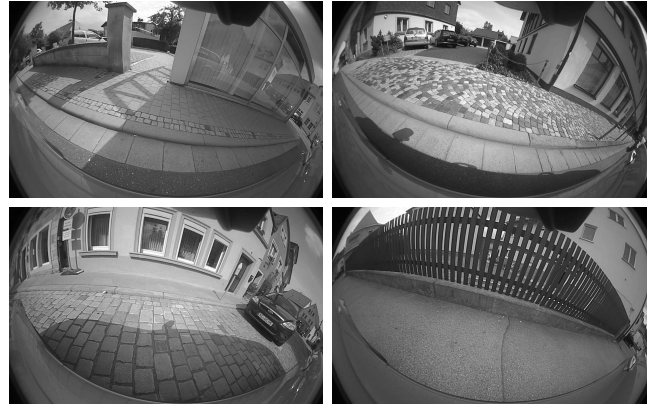}}
  %\centerline{}
%  \vspace{2.0cm}
\end{minipage}
%\hspace{-0.2cm}
\vspace{-0.7cm}
\caption{Example frames of fisheye videos incorporating radial distortion. From the top left to the bottom right: \textit{video1}, \textit{video2}, \textit{video3}, and \textit{video4}.}
\label{fig:exmp1}
\vspace{-0.5cm}
\end{figure}

This paper focuses on radial distortion and its effects on video coding.
It suggests a new processing order for the coding and distortion-correction of fisheye video material and also provides a comparison to the conventional processing chain.
The paper is organized as follows.
Section~\ref{sec:distcorr} deals briefly with 
%radial distortion and the correction thereof.
the acquisition of the parameters required for distortion correction.
In section~\ref{sec:methods}, the two different methods of coding and distortion correcting fisheye video sequences are described, the second of which is our proposed new processing order.
Section~\ref{sec:results} discusses the simulation setup and results, and section~\ref{sec:concl} gives a conclusion of this paper.

\section{Camera Calibration}
\label{sec:distcorr}

Due to the large field of view, images and videos captured by fisheye cameras exhibit strong radial distortions.
%The greater the distance from the image center, the greater the distortion gets, leading to strong deformations of the image content especially towards the periphery.
Fig.~\ref{fig:exmp1} shows some example frames of such distorted video material.
%The shown images represent example frames of the four video sequences used for the simulations presented later in this paper.
In many cases, it may be desirable to correct the distortion to obtain rectilinear image properties where straight lines in the scene actually appear as straight lines in the image -- just as if they were taken by a conventional perspective camera.
To that end, it is necessary to perform a camera calibration to determine the camera parameters.

For our purposes, we used a completely automatic technique that only requires a few calibration images showing a checkerboard pattern at different angles and positions~\cite{scara2}.
%The camera parameters are determined in a calibration step by using, for instance, calibration images that show a calibration pattern (e.\,g. a checkerboard pattern) at different angles and positions.
%This method makes use of a generalized omnidirectional camera model (also applicable to fisheye cameras) that assumes ...
Part of the calibration procedure is the estimation of the non-linear imaging function which effectively relates a point in the sensor plane to a scene point.
Part of this imaging function is described by the 4th-order polynomial
\begin{comment}
\vspace{-0.25cm}
\begin{equation}
g(u, v) = (u, v, f(u, v)^T)\,,
\end{equation}
with $u, v$ the sensor plane coordinates and 
\end{comment}
%The radial distortion can be described by the polynomial model
\begin{comment}
\begin{equation}
P = \begin{bmatrix}
x\\y\\z
\end{bmatrix}=\begin{bmatrix}
u\\v\\f(\rho)
\end{bmatrix}
\end{equation}
\end{comment}
\vspace{-0.1cm}
\begin{equation*}
f(u, v) = f(\rho) = a_0+a_1\rho+a_2\rho^2+a_3\rho^3+a_4\rho^4\,,
\vspace{-0.1cm}
\end{equation*}
with $\rho = \sqrt{u^2+v^2}$ and $u, v$ the sensor plane coordinates.
%$f(\rho)$ describes the mapping of a 3D vector of the scene onto the image plane.
$f(\rho)$ is a radially symmetric function of the Euclidean distance $\rho$ to the image center and the coefficients $a_n$ denote the intrinsic parameters, which the calibration procedure tries to approximate.

\begin{comment}
In how much detail should radial distortion be treated?
And the correction thereof?

\begin{itemize}
\item{distortion model (briefly; subsection?)}
\item{should provide equations (briefly)}
\item{cite Ocam Calib Toolbox}
\end{itemize}
\end{comment}

\begin{comment}
\begin{figure}[t]
\begin{minipage}[b]{0.5\linewidth}
  \centering
  \centerline{\includegraphics[width=4.0cm]{Figures/exmpFrame25_video1_420p_mod_DC-fc5}}
  %\centerline{}
%  \vspace{2.0cm}
\end{minipage}
\hspace{-0.2cm}
\begin{minipage}[b]{0.5\linewidth}
  \centering
  \centerline{\includegraphics[width=4.0cm]{Figures/exmpFrame25_video1_420p_mod_DC-fc9}}
  %\centerline{}
%  \vspace{2.0cm}
\end{minipage}
%
\vspace{-0.7cm}
\caption{One example frame of \textit{video1} after distortion correction with two different parameters $fc=5$ (left) and $fc=9$ (right).}
\label{fig:exmp1b}
%\vspace{-0.3cm}
\end{figure}
\end{comment}
\begin{figure}[t]
\begin{minipage}[b]{1.0\linewidth}
  \centering
  \centerline{\includegraphics[width=8.5cm]{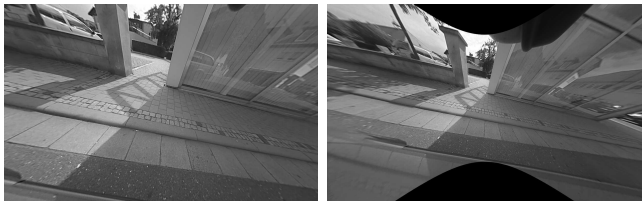}}
  %\centerline{}
%  \vspace{2.0cm}
\end{minipage}
\vspace{-0.7cm}
\caption{Rectified example frame of \textit{video1} using two different distance settings DCOR5 (left) and DCOR9 (right).}
\label{fig:exmp1b}
\vspace{-0.25cm}
\end{figure}
With the knowledge of the intrinsic parameters, the distortion can be corrected.
%The used toolbox allows choosing an estimation of the focal length $\mathit{fc}$ that enables an adjustment of the viewing distance.
The used toolbox allows adjusting the viewing distance via a zoom factor $\mathit{sf}$.
The greater this parameter is chosen, the higher the amount of image content that is preserved.
However, the perspective distortion becomes more pronounced.
Fig.~\ref{fig:exmp1b} shows a distortion-corrected example frame of \textit{video1} for two different distance settings, near (DCOR5) and far (DCOR9).

Further elaboration on the calibration procedure as well as the distortion correction is outside the scope of this paper.
Details are found in~\cite{scara2} and also in the tutorial accompanying the used toolbox~\cite{scara4}.
More general information on projective geometry and radial distortion and its correction may be found in~\cite{mvgeo,raddist,devernay}, for example.

% #####

\section{Methods for Distortion-Correction and Coding}
\label{sec:methods}

\begin{comment}
This is where the new processing order is proposed and explained.
Possibly shouldn't mix up state of the art and new proposal.
\hfill

Introduce the two methods and lead up to the two subsections.
\end{comment}

\begin{figure}[t]
\begin{minipage}[b]{1.0\linewidth}
  \centering
  \psfrag{a}[lB][lB][1.0][0]{\textit{postDCOR}}
  \psfrag{b}[lB][lB][1.0][0]{\textit{preDCOR}}
  \psfrag{c}[cc][lB][1.0][0]{DCOR}
  \psfrag{d}[cc][lB][1.0][0]{DEC}
  \psfrag{e}[cc][lB][1.0][0]{ENC}
  \centerline{\includegraphics[width=8.5cm]{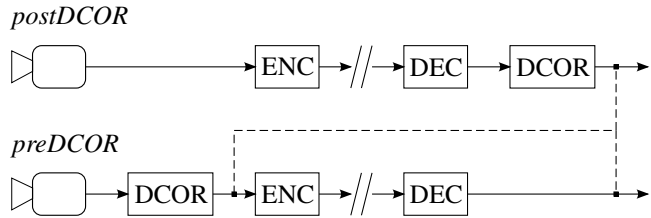}}
%  \vspace{2.0cm}
  %\centerline{Intra}\medskip
\end{minipage}
\vspace{-0.7cm}
\caption{State-of-the-art processing chain \textit{(postDCOR)} and proposed order \textit{(preDCOR)}. The dashed line indicates to which reference the output sequences are compared.}
\label{fig:procchain}
\vspace{-0.25cm}
\end{figure}

There are basically two possibilities to code and correct distorted videos.
The conventional way is to employ radial distortion correction (DCOR) as a post-processing step after decoding (\textit{postDCOR}).
The other possibility is to perform the correction first and then encode the corrected videos (\textit{pre\-DCOR}).
Fig.~\ref{fig:procchain} compares the two processing chains which are further described in the following.

\vspace{-0.5cm}
\subsubsection*{Post-correction of distorted video sequences}
\vspace{-0.2cm}

\begin{comment}
This is the state of the art.
Call it ``State of the art'' instead?

A very short statement about what is done so far (might be too brief for an own subsection):
\begin{itemize}
\item{Post-processing, application: cars, outdoor, etc.}
\item{no consideration for radial distortion}
\item{point out that at the end, there is a distortion corrected video!}
\item{MJPEG or H.264 coding; many cameras can do that}
\end{itemize}
\end{comment}

Typically, images and video sequences are captured by, for example, on-board car or outdoor cameras and directly encoded (e.\,g., in MotionJPEG or H.264/AVC~\cite{avc} format) for efficient storage.
The top half of Fig.~\ref{fig:procchain} depicts this state-of-the-art processing order.
If a user requires the data to be rectilinear, the correction is done after the data has been transmitted from the camera to some other device and decoded for further post-processing steps.
As the original, distorted data is encoded, the radial distortion of the images is not taken into consideration during compression, and the coding efficiency may deteriorate accordingly.

\vspace{-0.5cm}
\subsubsection*{Distortion correction as a pre-processing step}
\vspace{-0.2cm}

\begin{comment}
Here, the new scheme is introduced.
\end{comment}

The bottom half of Fig.~\ref{fig:procchain} shows our proposed processing order.
In contrast to the post-correction case, we propose the distortion correction to be done as a pre-processing step before encoding the captured data.
That way, the images and videos to be coded are rectilinear and thus perfectly suited for both temporal block-matching and spatial angular prediction.
Of course, this processing order requires the camera parameters to be already available, i.\,e., the calibration has to be done beforehand as well.
As there are no deteriorating effects caused by radial distortion, the coding efficiency is expected to increase for this pre-correction method.
This is shown to be true in the next section.

\begin{comment}
Thus, the captured video frames are only encoded after the radial distortion has been removed.
The encoded bitstreams are again transmitted or stored, and decoded at the receiver side for further usage.
\end{comment}

%#####

\section{Simulation Results}
\label{sec:results}
\vspace{-0.3cm}

\begin{comment}
\begin{figure}[h]
\begin{minipage}[b]{1.0\linewidth}
  \centering
  \centerline{\includegraphics[width=8.0cm]{Figures/prepostPSNRv1I24-40}}
%  \vspace{2.0cm}
  \centerline{Intra}\medskip
\end{minipage}
\begin{minipage}[b]{1.0\linewidth}
  \centering
  \centerline{\includegraphics[width=8.0cm]{Figures/prepostPSNRv1LD24-40}}
%  \vspace{2.0cm}
  \centerline{Low Delay}\medskip
\end{minipage}
\begin{minipage}[b]{1.0\linewidth}
  \centering
  \centerline{\includegraphics[width=8.0cm]{Figures/prepostPSNRv1RA24-40}}
%  \vspace{2.0cm}
  \centerline{Random Access}\medskip
\end{minipage}
\vspace{-1.0cm}
%
\caption{Rate-distortion curves for \textit{video1}.}
\label{fig:vid1b}
\vspace{-0.3cm}
\end{figure}
\end{comment}
\begin{comment}
\begin{figure*}[ht]
\begin{minipage}[b]{0.33\linewidth}
  \centering
  \centerline{\includegraphics[width=5.5cm]{Figures/prepostPSNRv1I24-40}}
%  \vspace{2.0cm}
  \centerline{Intra}\medskip
\end{minipage}
\begin{minipage}[b]{0.33\linewidth}
  \centering
  \centerline{\includegraphics[width=5.5cm]{Figures/prepostPSNRv1LD24-40}}
%  \vspace{2.0cm}
  \centerline{Low Delay}\medskip
\end{minipage}
\begin{minipage}[b]{0.33\linewidth}
  \centering
  \centerline{\includegraphics[width=5.5cm]{Figures/prepostPSNRv1RA24-40}}
%  \vspace{2.0cm}
  \centerline{Random Access}\medskip
\end{minipage}
\vspace{-1.0cm}
%
\caption{Rate-distortion curves for \textit{video1}.}
\label{fig:vid1b}
\vspace{-0.3cm}
\end{figure*}
\end{comment}

\begin{figure*}[ht]
\begin{minipage}[b]{1.0\linewidth}
  \centering
  \psfrag{Intra}[cc][cc][1.0][0]{Intra}
  \psfrag{Low Delay}[cc][cc][1.0][0]{Low Delay}
  \psfrag{Random Access}[cc][cc][1.0][0]{Random Access}
  \psfrag{Rate in bpp}[ct][cc][1.0][0]{Rate in bpp}
  \psfrag{PSNR Y in dB}[cB][cc][1.0][0]{PSNR Y in dB}
  \psfrag{31}[cc][cc][1.0][0]{31}
  \psfrag{33}[cc][cc][1.0][0]{33}
  \psfrag{35}[cc][cc][1.0][0]{35}
  \psfrag{37}[cc][cc][1.0][0]{37}
  \psfrag{39}[cc][cc][1.0][0]{39}
  \psfrag{41}[cc][cc][1.0][0]{41}
  \psfrag{43}[cc][cc][1.0][0]{43}
  \psfrag{45}[cc][cc][1.0][0]{45}
  \psfrag{0}[cc][cc][1.0][0]{0}
  \psfrag{0.1}[cc][cc][1.0][0]{0.1}
  \psfrag{0.2}[cc][cc][1.0][0]{0.2}
  \psfrag{0.3}[cc][cc][1.0][0]{0.3}
  \psfrag{0.4}[cc][cc][1.0][0]{0.4}
  \psfrag{0.5}[cc][cc][1.0][0]{0.5}
  \psfrag{0.05}[cc][cc][1.0][0]{0.05}
  \psfrag{0.15}[cc][cc][1.0][0]{0.15}
  \psfrag{preDCOR5}[cl][cl][0.8][0]{preDCOR5}
  \psfrag{postDCOR5}[cl][cl][0.8][0]{postDCOR5}
    \psfrag{preDCOR7}[cl][cl][0.8][0]{preDCOR7}
    \psfrag{postDCOR7}[cl][cl][0.8][0]{postDCOR7}
      \psfrag{preDCOR9}[cl][cl][0.8][0]{preDCOR9}
      \psfrag{postDCOR9}[cl][cl][0.8][0]{postDCOR9}
  \centerline{\includegraphics[width=21cm]{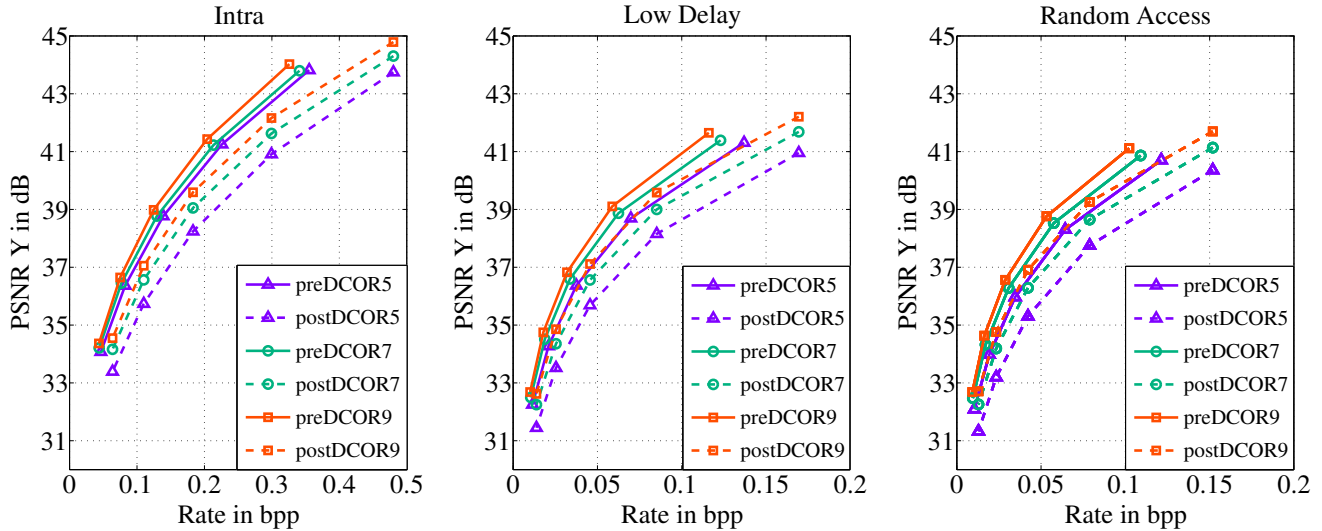}}
  \vspace{0.3cm}
%  \centerline{Intra}\medskip
\end{minipage}
\vspace{-1.0cm}
\caption{Rate-distortion curves for \textit{video1}.}
\label{fig:vid1b}
\vspace{-0.3cm}
\end{figure*}

\begin{table*}[t]
\caption{$\Delta$PSNR in dB for the four test sequences.}
\label{tab:bdpsnr1}
\vspace{0.1cm}
\centering
\renewcommand\arraystretch{0.8}
\begin{tabularx}{\textwidth}{p{1.0cm}p{0cm}cccp{0.3cm}cccp{0.3cm}ccc} % textwidth for page width
\toprule
& & \multicolumn{3}{c}{\textbf{Intra}} & & \multicolumn{3}{c}{\textbf{Low Delay}} & & \multicolumn{3}{c}{\textbf{Random Access}} \\
 & & DCOR5 & DCOR7 & DCOR9 & & DCOR5 & DCOR7 & DCOR9 & & DCOR5 & DCOR7 & DCOR9 \\
\midrule
\textit{video1} & & 1.91 & 1.44 & 1.33 & & 1.37 & 1.13 & 1.06 & & 1.37 & 1.11 & 1.02 \\ 
\textit{video2} & & 1.16 & 1.01 & 1.44 & & 0.89 & 0.99 & 1.39 & & 1.01 & 1.04 & 1.37 \\
\textit{video3} & & 1.84 & 1.09 & 1.11 & & 1.19 & 0.80 & 0.79 & & 1.19 & 0.77 & 0.76 \\
\textit{video4} & & 0.68 & 0.41 & 0.56 & & 0.65 & 0.48 & 0.47 & & 0.73 & 0.60 & 0.60 \\
\addlinespace
mean & & 1.40 & 0.99 & 1.11 & & 1.03 & 0.85 & 0.92 & & 1.07 & 0.88 & 0.94 \\
\bottomrule
\end{tabularx}
\end{table*}

For our simulations, we tested four traffic video sequences captured by fisheye car cameras.
Each sequence comprises 30 frames and only the luminance component of the raw YUV sequences was considered.
Example frames are depicted in Fig.~\ref{fig:exmp1}.
For camera calibration and distortion correction, the \textit{OCamCalibToolbox} by Davide Scaramuzza~\cite{scara2, scara4} was used.
The distortion-corrected sequences were generated for three different viewing distances: near, medium far, and far, i.\,e., $\mathit{sf}\in\{5,7,9\}$, also denoted as DCOR5, DCOR7, and DCOR9 (for a visual example, cf.~Fig.~\ref{fig:exmp1b}).
As a hybrid video codec, the reference software HM-11.0~\cite{refsoft} of H.265/HEVC was selected.
The three Main profile configurations Intra, Low Delay, and Random Access were tested for five quantization parameters ranging between 24 and 40.
The following two subsections discuss the results of the rate-distortion analysis and perceptual quality evaluation, respectively.
\begin{comment}
\begin{figure*}[ht]
\begin{minipage}[b]{0.5\linewidth}
  \centering
  \centerline{\includegraphics[width=8.5cm]{Figures/prepostPSNRv1I24-40}}
%  \vspace{2.0cm}
  \centerline{Intra}\medskip
\end{minipage}
\begin{minipage}[b]{0.5\linewidth}
  \centering
  \centerline{\includegraphics[width=8.5cm]{Figures/prepostPSNRv1RA24-40}}
%  \vspace{2.0cm}
  \centerline{Random Access}\medskip
\end{minipage}
\vspace{-1.0cm}
%
\caption{Rate-distortion curves for \textit{video1} [TODO, quality of plot].}
\label{fig:vid1b}
\vspace{-0.3cm}
\end{figure*}
\end{comment}
\begin{comment}
\begin{itemize}
\item{4 videos, 30 frames each (corresponds to 1 or 2 seconds depending on video)}
\item{HEVC HM-11.0, Intra, Low Delay, Random Access modes, Main profile}
\item{QPs ranging from 24 through 40 (or 16 if more data points are used)}
\item{reference is the distortion correction version of each video}
\item{to be independent of the framerate, bits per pixel are used for the rate}
\item{for the PSNR, only the luminance PSNR is considered}
\item{in general, only the luminance component is considered when processing the videos}
\item{for the distortion correction, three different parameter settings for ``fc'' were used}
\item{Fig.~\ref{fig:exmp1b} shows the two example frames depicted above after distortion correction has been employed.}
\item{Fig.~\ref{fig:exmp1c} shows an alternative version of Fig.~\ref{fig:exmp1b}}
\end{itemize}
\end{comment}

\begin{table}[t]
\caption{Average bitrate differences in \% based on PSNR.}
\label{tab:bdpsnr2}
\vspace{0.1cm}
\centering
\renewcommand\arraystretch{0.8}
\begin{tabularx}{\linewidth}{p{3cm}ccc} % linewidth for column width
\toprule
 & DCOR5 & DCOR7 & DCOR9 \\
\midrule
Intra & -22.07 & -16.68 & -18.59 \\
Low Delay & -21.79 & -18.94 & -20.35 \\
Random Access & -23.25 & -20.13 & -21.29 \\ 
\bottomrule
\end{tabularx}
\vspace{-0.5cm}
\end{table}

\begin{table*}[t]
\caption{$\Delta$SSIM results (scaled by a factor of 100) for the four test sequences.}
\label{tab:ssim}
\vspace{0.1cm}
\centering
\renewcommand\arraystretch{0.8}
\begin{tabularx}{\textwidth}{p{1cm}p{0cm}cccp{0.3cm}cccp{0.3cm}ccc}
\toprule
& & \multicolumn{3}{c}{\textbf{Intra}} & & \multicolumn{3}{c}{\textbf{Low Delay}} & & \multicolumn{3}{c}{\textbf{Random Access}} \\
 & & DCOR5 & DCOR7 & DCOR9 & & DCOR5 & DCOR7 & DCOR9 & & DCOR5 & DCOR7 & DCOR9 \\
\midrule
\textit{video1} & & 2.23 & 1.43 & 1.16 & & 2.21 & 1.59 & 1.27 & & 2.36 & 1.61 & 1.26 \\
\textit{video2} & & 1.25 & 1.08 & 1.22 & & 1.57 & 1.55 & 1.71 & & 1.80 & 1.71 & 1.85 \\
\textit{video3} & & 1.66 & 1.06 & 1.08 & & 1.80 & 1.29 & 1.26 & & 1.94 & 1.34 & 1.29 \\
\textit{video4} & & 1.82 & 1.41 & 1.14 & & 2.38 & 1.83 & 1.30 & & 2.69 & 2.05 & 1.45 \\
\addlinespace
mean & & 1.74 & 1.25 & 1.15 & & 1.99 & 1.57 & 1.39 & & 2.20 & 1.68 & 1.46 \\
\bottomrule
\end{tabularx}
\end{table*}

%\vspace{0.1cm}
\subsection{Rate-Distortion Analysis}
%\vspace{-0.1cm}

In terms of objective video quality, the luminance PSNR (PSNR Y) was evaluated for different quantization parameters.
For this purpose, the output sequences of both the \textit{postDCOR} and the \textit{preDCOR} processing chain were compared to the distortion-corrected versions of the original video sequences as indicated by the dashed line in Fig.~\ref{fig:procchain}.
Furthermore, the bitrate in bits per pixel (bpp) was determined.
Fig.~\ref{fig:vid1b} shows indicative rate-distortion results for one of the four test sequences.
%On the left, the curves for the Intra mode are shown, in the middle, those for the Low Delay mode, and on the right, the results for the Random Access mode are depicted.
From left to right, the curves for the Intra, Low Delay, and Random Access mode are depicted, respectively.
%With the exception of slightly higher bitrates, the results for the Low Delay mode are similar to those of the Random Access mode so that the latter has been chosen as a representative for inter-frame coding.
The results for the remaining three video sequences are similar to the ones shown.

\begin{comment}
\begin{figure*}[t]
\begin{minipage}[b]{0.5\linewidth}
  \centering
  \centerline{\includegraphics[width=8.5cm]{Figures/prepostPSNRv2I24-40}}
%  \vspace{2.0cm}
  \centerline{Intra}\medskip
\end{minipage}
\begin{minipage}[b]{0.5\linewidth}
  \centering
  \centerline{\includegraphics[width=8.5cm]{Figures/prepostPSNRv2RA24-40}}
%  \vspace{2.0cm}
  \centerline{Random Access}\medskip
\end{minipage}
\vspace{-1.0cm}
%
\caption{Rate-distortion curves for \textit{video2}.}
\label{fig:vid2b}
\vspace{-0.3cm}
\end{figure*}
\end{comment}
Comparing each dashed line to the corresponding solid one yields a PSNR Y increase of up to 2 dB, corresponding to bitrate savings of about 10 to 20~\%.
This holds true for all three distortion-corrected versions of the sequences.
To create average PSNR results over all tested quantization parameters, the Bj\o{}ntegaard quality metric (BD-PSNR)~\cite{bdpsnr} was employed.
Table~\ref{tab:bdpsnr1} provides the BD-PSNR results in the form of PSNR differences for all four video sequences.
%The $\Delta$PSNR values were obtained by subtracting the \textit{postDCOR} PSNR results from the \textit{preDCOR} PSNR results.
The $\Delta$PSNR values were obtained using the \textit{postDCOR} PSNR curves as reference curves.
Positive values hence denote a gain, meaning that a quality improvement is achieved when using the \textit{preDCOR} chain.
Obviously, gains are achieved throughout all tests with an overall maximum of 1.91 dB and average gains of 1.17 dB for intra-frame and 0.95 dB for inter-frame coding.
Equivalent maximum bitrate savings amount to 31.76~\%, while on average, 19.11~\% and 20.96~\% can be saved for intra-frame and inter-frame coding, respectively.
Table~\ref{tab:bdpsnr2} summarizes the average relative bitrate differences for the three modes.
Negative values denote a bitrate reduction, so that again, the \textit{preDCOR} method always performs better than the \textit{postDCOR} method for the tested sequences.

\vspace{-0.2cm}
\subsection{Perceptual Quality Evaluation}
%\vspace{-0.2cm}

\begin{figure}[t]
\begin{minipage}[b]{1.0\linewidth}
  \centering
  \centerline{\includegraphics[width=8.5cm]{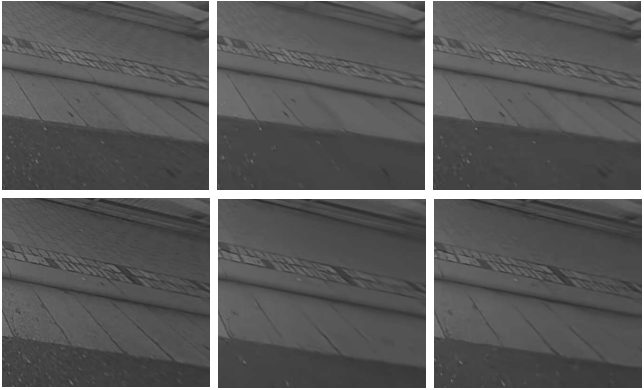}}
  %\centerline{}
%  \vspace{2.0cm}
\end{minipage}
%\hspace{-0.2cm}
\vspace{-0.7cm}
\caption{Example of the visual quality improvement for two frames of \textit{video1}. Top row: frame 10, Intra mode. Bottom row: frame 28, Random Access mode. From left to right: reference frame, \textit{postDCOR} method, \textit{preDCOR} method.}
\label{fig:visualcomparison}
\vspace{-0.3cm}
\end{figure}

For further evaluation and comparison of the visual quality, the structural similarity (SSIM) index~\cite{ssim1} was used as a perceptual quality metric.
%For all results it was true that the higher the quantization, the lower the SSIM value. 
%Example: 0.99XX decreases down to 0.90XX~--~0.85XX.
The distortion-corrected original videos served again as reference sequences for both the \textit{postDCOR} and the \textit{preDCOR} chain.
In order to provide a fair comparison, the SSIM evaluation was done equivalently to the rate-distortion analysis, i.\,e., the averaging is also based on the Bj\o{}ntegaard method.
%Like before, differences were calculated to determine whether a quality gain or loss is achieved.
Like before, the \textit{postDCOR} results served as the reference results to determine whether a quality gain or loss is achieved.
Table~\ref{tab:ssim} contains the resulting $\Delta$SSIM values (scaled by a factor of 100) that were obtained for each sequence and each mode.
Positive values denote a gain, so that it can be observed that the proposed \textit{preDCOR} processing order manages to improve the visual quality of the sequences throughout all tests conducted.
%For lower quantization parameters, the difference is much less (e.\,g., 0.001 points), but nevertheless, the SSIM values are always higher for the \textit{preDCOR} method, implying that a better visual quality can be achieved for the same amount of quantization.
Average gains amount to 1.38 points for intra-frame and 1.71 points for inter-frame coding.
For the actual SSIM values, note that a minimum of 0.803 for coarse quantization and a maximum of 0.996 for fine quantization was observed.

In addition to the above, a visual comparison of the two output sequences was made to confirm the visual quality improvements implied by the PSNR and SSIM results.
Fig.~\ref{fig:visualcomparison} shows an example for intra-frame (top) and inter-frame coding (bottom) using a quantization parameter of 32.
The reference frame (DCOR5, left) is compared to the two processing methods (middle and right).
As can be seen, straight lines are better reconstructed with the \textit{preDCOR} method.
%It was also observed that the original structures are better retained.
%Thus, the subjective results are in favor of the objective ones.
%[TODO could provide a (short!) table with examples; limit to a few average values if so]

\section{Conclusion}
\label{sec:concl}

\begin{comment}
Sum up again what is done and what is proposed. 
What are the results and what do they tell us?
If distortion correction is done as a post-processing step anyway (application-dependent),
then it makes sense to do it as a pre-processing step instead as this leads to more efficient coding and thus storing.
Gains in both bitrate (i.\,e., bitrate savings) and luminance PSNR are achieved both in Intra and Inter (LD and RA) mode.

Outlook in one sentence: So that it can be concluded that radial distortion correction could prove beneficial if incorporated into a video coder as an alternative prediction mode for both intra and inter coding.
\end{comment}

In this paper, we compared two different processing orders for the coding and distortion correction of radially distorted video sequences.
The conventional order that employs distortion correction as a post-processing step was compared to our proposed order of employing distortion correction as a pre-processing step. 
Evaluation results showed that coding gains as well as visual quality improvements are achieved for the proposed method throughout all tests conducted.
We conclude that radial distortion and compensation thereof is a promising means for improving the coding efficiency of hybrid video codecs such as H.265/HEVC.

As there exists no prior work in that regard, future work will include further investigation and analysis of the influence of radial distortion on intra-frame and inter-frame prediction in hybrid video coding and make an effort to exploit the distortion properties for more efficient coding techniques.
Similar to the rotational motion estimation described in~\cite{springer}, radial distortion correction could be incorporated into H.265/HEVC as a compensation procedure to that end.

%#####

\section{Acknowledgment}
\vspace{-0.15cm}

This work was partly supported by the Research Training Group 1773 “Heterogeneous Image Systems”, funded by the German Research Foundation (DFG).
The original video sequences and the calibration results have been kindly provided by Continental Chassis \& Safety  BU ADAS Segment Surround View (A.D.C GmbH), Kronach.

%\vfill
%\pagebreak

%\newpage

%\section{TODO}
%\label{sec:ref}

%formatting, plots (axes, font, position on page, etc.), tables (SSIM examples, shorten BD-PSNR?), remove last page, rest must fit on first 4 pages!, refs

% BiBTeX files (here: refs). The IEEEbib.bst bibliography
\bibliographystyle{IEEEbib}
\bibliography{refs}

\begin{comment}
%delete the following lines once the ref/bib is done (just a placeholder!)
\section{References}

\begin{itemize}
\item{HEVC standard document}
\item{HM URL}
\item{Toolbox papers}
\item{Toolbox URL}
\item{Tech. report?}
\item{Bjontegaard VCEG-M33}
\item{SSIM if used}
\item{radial distortion correction (model)}
\end{itemize}
\end{comment}

\end{document}